\documentclass[10pt,conference]{IEEEtran}
\IEEEoverridecommandlockouts
\usepackage{cite}
\usepackage{amsmath,amssymb,amsfonts}
\usepackage{algorithmic}
\usepackage{graphicx}
\usepackage{subcaption}
\usepackage{textcomp}
\usepackage{xcolor}
\usepackage{hyperref}
\usepackage{xurl}
\usepackage{booktabs}
\usepackage{tabularx}
\usepackage{threeparttable}
\usepackage{multirow}
\usepackage{dblfloatfix}
\usepackage{caption}
\usepackage{booktabs}
\usepackage{subcaption}

\usepackage{tcolorbox}

\usepackage{cite}
\usepackage{amsmath,amssymb,amsfonts}
\usepackage{algorithmic}
\usepackage{graphicx}
\usepackage{textcomp}
\usepackage{booktabs} 
\usepackage{xcolor}
\usepackage{scalerel}
\usepackage{hyperref}
\usepackage{orcidlink}
\hypersetup{
    colorlinks=true,
    linkcolor=black,
    filecolor=black,      
    urlcolor=black,
    citecolor=blue,
    pdftitle={Evaluating Network Impact of Post-Quantum Certificate Chain sizes on Time to First Byte in TLS Deployments},
    pdfpagemode=FullScreen,
    }
\def\BibTeX{{\rm B\kern-.05em{\sc i\kern-.025em b}\kern-.08em
    T\kern-.1667em\lower.7ex\hbox{E}\kern-.125emX}}
\begin{document}

\title{Network Impact of Post-Quantum Certificate Chain sizes on Time to First Byte in TLS Deployments \\
}

\author{\IEEEauthorblockN{Matthew Chou \orcidlink{0009-0009-4506-1142}$^1$, Phuong Cao\thanks{*Corresponding author: Phuong Cao; Data: \href{https://pmcao.github.io}{https://pmcao.github.io}} \orcidlink{0000-0001-6028-0583}$^{1,2,*}$
\\}
\IEEEauthorblockA{
\small{
$^1$University of Illinois at Urbana-Champaign,
$^2$National Center for Supercomputing Applications
}}
}

\maketitle

\begin{abstract}
Post-Quantum Cryptography (PQC) is a rapidly growing deployment challenge as cryptographically relevant quantum computers (CRQC) continue to advance, leaving traditional cryptographic algorithms used in X.509 vulnerable to attack. However, PQC introduces significant deployment challenges in real-world networks, with handshake sizes increasing from 5x to over 20x compared to classical algorithms. In this work, we evaluate the time to first byte (TTFB) under CDN-focused TLS conditions to characterize the latency cost of transitioning existing internet infrastructure to quantum-safe certificate schemes. We observe discrete increases in TTFB as certificate chain sizes exceed transport layer data flight limits. To isolate the impact of certificate chains, we evaluate both ECDSA and ML-DSA-based certificate schemes, generating similarly sized certificate chains through controlled addition of certificate extensions. We additionally examine how CDN properties such as session resumption, certificate size optimizations, and geographical distribution reduce latency penalties. To ground our findings in real-world applications, we utilize Zeek-monitored TLS traffic through a High-Performance Computing System (NCSA) with terabyte network connectivity across the nation to quantify real-world session resumption rates. We compare CDN-driven size optimization with Merkle Tree Certificates (MTC) to examine how size reductions allow certificate chains to remain under the flight limit threshold. We find that MTC allows for 2x-3x increase in supportable certificate chain size, whereas CDN-based optimizations yield more limited reductions, supporting up to approximately 1.6x certificate chain size increase. These findings reveal a bandwidth-rooted bottleneck in PQC deployment latency, highlighting the critical importance of certificate chain design and optimization strategies for achieving a practical quantum-safe internet infrastructure.
\end{abstract}


\section{Introduction}
Cryptographically relevant quantum computers (CRQCs) are expected to break real-world encryption schemes using Shor's algorithm, also known as Q-Day \cite{shor1994}. Although quantum computers are still years away from being strong enough to break modern encryption algorithms such as RSA, there remains the threat of attackers who will harvest encrypted data now to be decrypted later when quantum computers become available. This motivates the development of Post-Quantum Cryptography (PQC), i.e., algorithms resistant to quantum computing \cite{mosca2018cybersecurity}. The National Institute of Standards and Technology is standardizing PQC and developing deployable quantum-resistant algorithms for key exchange/encryption and digital signatures \cite{nist_pqc}.

This work evaluates how certificate chain length directly affects TTFB, specifically using both traditional and Post-Quantum-based certificate schemes and artificially inflating their sizes to isolate the network-derived latency. We vary the round-trip time (RTT), the number of intermediate certificates, and the certificate size to further analyze the effects of end-to-end delays. We observe that the dominant factor contributing to TTFB is not propagation delay but rather bandwidth limitations and implementation overhead. Our work also proposes using CDN-specific behaviors as well as Merkle Tree Certificates (MTC)\cite{ietf_mtc_2026} to reduce RTT by keeping certificate chain sizes within the flight capacity threshold.

However, while these algorithms do protect against quantum adversaries, they have several drawbacks. PQC secures data using mathematically complex problems, such as lattice problems and high-dimensional algebra, thereby substantially increasing the number of bits required for both keys and certificates. Certificate sizes, in particular, increase substantially, ranging from 5x to over 20x depending on the specific scheme used \cite{alagic2022nist8413,sim2025kpqc}. This negatively impacts Transport Layer Security (TLS) handshakes as packets exceed the Maximum Transmission Unit (MTU), leading to data fragmentation and increased round-trip times (RTT). These directly affect the time to first byte (TTFB), a user-facing performance metric\cite{rfc8446}.

While the TLS protocol remains consistent across deployment environments, higher latencies affect infrastructure differently. Content Delivery Networks, which are built to minimize latency by serving content from geographically adjacent edge servers, represent a distinct case compared to traditional non-CDN origin-based systems\cite{nygren2010akamai}. The increased latency introduced by PQC would significantly affect CDN infrastructure, as these latency-optimized environments handle millions of connections where even small per-connection delay increases can translate into substantial consequences at scale\cite{cloudflare_pq_tls_2019}.

There have been many studies evaluating PQC in TLS environments on handshake latencies and PQC overhead\cite{barton2019pqc_tls,bos2015postquantum}. However, these works do not isolate the effect of certificate chain sizes on TTFB, nor do they explicitly model transport layer flight limits. We use data from the Zeek measurement tool \cite{zeek_ssl_main,paxson1998bro} with NCSA's database to analyze CDN-specific properties such as session resumption rates, providing real-world conditions for our experiments. 

\textbf{Results.} Merkle Tree Certificates and CDN chain optimizations allow for certificate chain size increases by $\sim$2x-3x and up to $\sim$1.6x respectively while still keeping certificate chain size within bandwidth control window thresholds. Moreover, session resumption is able to keep certificates under bandwidth limits, with CDNs saving 2x more time to first byte (TTFB) than non-CDNs on average. Our data is grounded in real results with us seeing $\sim$80\%-90\% Session Resumption rates for CDNs. We additionally observed CDN session resumption rates are proportinal to TLS 1.3 adoptions rates over our 16 month dataset.
\newline
\textbf{Contributions.} A summary of our main contributions is as follows:
\begin{itemize}
    \item Analysis of the effect of certificate chain size on TTFB across both ECDSA and ML-DSA-based schemes, allowing for direct evaluation of TLS latencies due to propagation delay.
    \item Comparison of CDN impact on PQC including 1) geographical distribution, 2) certificate optimization, and 3) session resumption.
    \item Evaluation of MTC and CDN optimizations as chain size reduction techniques, comparing their effectiveness in reducing overhead by enabling certificate chains to remain within packet flight limits.
    \item Zeek-monitored TLS traffic of an NCSA to analyze real-world session resumption behavior.
\end{itemize}
\textbf{Putting this Paper in Perspective.} Prior works have studied handshake size on network latency for various PQC certificates. Numerous other works have analyzed CDN optimizations, implemented with real data. However, none of the previous works have directly compared the time to first byte (TTFB), while isolating certificate chain sizes across various CDN and MTC optimizations.

\begin{table*}
\centering
\begin{threeparttable}
\caption{Our Comparison of Certificate Schemes and Structures and Their Impact on TLS\cite{alagic2022nist8413,bos2015postquantum,ietf_mtc_2026}}
\label{tab:pqc_comparison}
\renewcommand{\arraystretch}{1.2}
\begin{tabularx}{\textwidth}{l l X X X l}
\toprule
\textbf{Scheme} & \textbf{Type} & \textbf{Cert Size} & \textbf{Chain Size} & \textbf{Handshake Impact} & \textbf{Status} \\
\midrule
RSA        & Classical (integer factorization) & Small ($\sim$1--2 KB) & Small ($\sim$3--5 KB) & Low latency & Deployed \\
ECDSA      & Classical (elliptic curve) & Small ($\sim$0.5--1.5 KB) & Small ($\sim$2--4 KB) & Low latency & Widely deployed \\
ML-DSA     & PQC (lattice) & Large ($\sim$2.5--4 KB) & Large ($\sim$8--15 KB) & Increased latency & Emerging \\
FALCON     & PQC (lattice) & Moderate ($\sim$1--2 KB) & Moderate ($\sim$5--10 KB) & Moderate overhead & Experimental \\
Hybrid ML-DSA & Hybrid (classical + PQC) & Very large (($\sim$3.5--6 KB)) & Very large ($\sim$12--25 KB) & High overhead & Transition \\
SLH-DSA    & PQC (hash-based) & Very large ($\sim$20--40 KB) & Very large ($\sim$60--150 KB) & High overhead & Experimental \\
ML-DSA + MTC & Structure (Merkle-based) &  $\sim$2.5--4 KB (logical), $\sim$0.7--1.0 KB (proof)\tnote{a} & Same as cert size & Reduced handshake size & Research \\
SLH-DSA + MTC & Structure (Merkle-based) &  $\sim$2.5--4 KB (logical), $\sim$0.7--1.0 KB (proof)\tnote{a} & Same as cert size & Reduced handshake size & Research \\
\bottomrule
\end{tabularx}
\begin{tablenotes}
\footnotesize
\item[a] Assuming a Merkle tree with $N \approx 2^{24}$ - $2^{28}$ leaves and SHA-256 hashing, yielding proof sizes as demonstrated in Section~\ref{subsec:merkle}.
\end{tablenotes}
\end{threeparttable}
\end{table*}

\section{Background}
This section provides background on modern web security and post-quantum cryptography (PQC). It covers TLS fundamentals, post-quantum cryptography (PQC) basics, relevant certificate types, and key CDN system features.

\subsection{Transport Layer Security Overview}
Transport Layer Security (TLS) is a cryptographic protocol designed to provide secure communication over digital networks. It is most commonly used to securely request and send web traffic through HTTPS and builds a foundation for modern Internet security. TLS ensures confidentiality through the use of key exchanges, digital signatures, and symmetric encryption\cite{rfc8446}.

TLS establishes a secure channel via a handshake that negotiates the protocol version, cipher suite, certificate, and key exchange method before initiating symmetric encryption. An important step of the handshake relevant to this paper is digital signature verification, in which a server proves its possession of a private key corresponding to the public key in its certificate. Authentication of certificates in TLS is achieved through public key infrastructure (PKI), requiring servers to verify their digital signature. Certificate chains provide a hierarchy of trust for the browser to verify a server's identity. There exist three levels of certificates: leaf, intermediate, and root, with each certificate being signed by the level above it and the root certificate serving as a trust anchor that is pre-installed in the client's device\cite{rfc5280}.

Cipher Suites and protocol versions provide guidelines on how information is sent and interpreted. There exist two main TLS protocols, namely TLS 1.2 and TLS 1.3, with TLS 1.3 reducing latency, providing additional security, and allowing for easier implementation of PQC. However, not all systems have updated to TLS 1.3 due to legacy architectures and difficulty of deployment\cite{rfc8446}.

TLS handshake delay stems from multiple sources. Oversized packets exceeding the MTU are segmented, increasing the chance of drops and retransmissions \cite{rfc8446} and thus adding time to the handshake. TCP also imposes bandwidth constraints: the initial congestion window (IW) caps first-RTT transmission at ~14KB, requiring an extra RTT if exceeded \cite{rfc6928}, and TCP slow start doubles the window each RTT, creating an effective secondary threshold of ~28KB, so transmissions exceeding ~42KB total incur yet another RTT.

\subsection{Post Quantum Cryptography Fundamentals}
PQC was developed in response to the threat of cryptographically relevant quantum computers (CRQCs), which could break asymmetric algorithms like RSA and ECDHE via Shor's algorithm and weaken symmetric and hash functions via Grover's algorithms \cite{shor1994,grover1996database}. The most pressing current threat is "harvest now, decrypt later," where adversaries collect encrypted data today to decrypt once CRQCs mature, potentially within 10–20 years \cite{mosca2018cybersecurity}, enabling compromise of key exchange and possible signature forgery.

In TLS, PQC augments or replaces classical algorithms for key exchange and digital signatures. The most widely adopted algorithms are ML-KEM (CRYSTALS-Kyber) for key exchange and ML-DSA (CRYSTALS-Dilithium) for signatures \cite{fips203}. Since these algorithms are not yet fully optimized for real-world deployment, hybrid implementations are common, combining PQC with traditional algorithms to maintain security against both quantum and classical adversaries \cite{stebila2025hybridtls}. 

PQC key exchange, led by ML-KEM, has seen widespread adoption, while PQC certificates remain largely experimental due to their larger sizes and PKI integration requirements. This proves to be a challenge, as the certificate chain structure typically requires 1-2 additional certificates as intermediates, drastically increasing total handshake sizes. These larger certificate sizes make handshakes more sensitive to congestion window limits, and as a result, real-world deployments must balance security and performance, specifically in latency-sensitive structures such as CDNs\cite{nygren2010akamai}.

\begin{figure*}[t]
    \centering
    \includegraphics[width=1.0\textwidth]{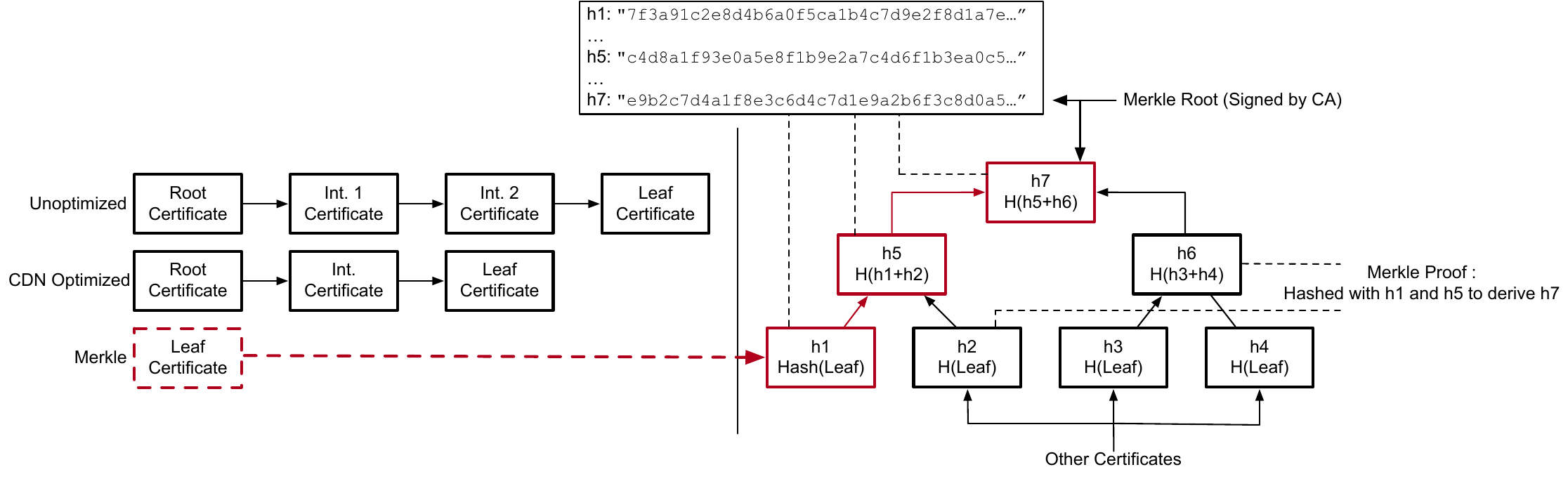}
    \caption{Merkle Based Certificate Structure Compared to Traditional X.509 Structure}
    \label{fig:merkle}
\end{figure*}

\subsection{Certificate Schemes in Classical and Post-Quantum TLS}
Many different types of certificates exist, with more being introduced as the field of PQC grows. This section provides an overview of common certificates and outlines the certificate schemes evaluated in this work. We provide data comparisons in Table~\ref{tab:pqc_comparison}. We ultimately settled on simulating ECDSA, ML-DSA, SLH-DSA, and Merkle Trees configurations. 

\subsubsection{Traditional}
The two most common traditional certificates are Rivest–Shamir–Adleman (RSA) \cite{rivest1978rsa} and Elliptic Curve Digital Signature Algorithm (ECDSA) \cite{rfc8422}, with RSA being more common in older systems. RSA uses the mathematical hardness of a prime factoring problem, often containing larger certificate sizes than the elliptic curve-based ECDSA certificates. ECDSA is increasingly preferred with 2-3x smaller certificates, lowering bandwidth and providing faster signatures, but sometimes being incompatible with legacy systems\cite{durumeric2013https}.

\subsubsection{Post-Quantum}
Given that traditional certificates are being broken by Shor's algorithm, newer certificates require different mathematically hard problems, the most common being lattice or hash-based. The most common digital signatures are Module-Lattice-Based Digital Signature Algorithm (ML-DSA), formerly known as CRYSTALS-Dilithium, and Stateless Hash-Based Digital Signature Algorithm (SLH-DSA), which is based on SPHINCS+. NIST has approved both algorithms, with ML-DSA being the primary standard due to its balance of security, performance, and size\cite{fips204}. SLH-DSA, while standardized, remains mostly impractical due to its huge certificate sizes\cite{alagic2022nist8413}. There also exist other algorithms, such as Fast Fourier Lattice-based Compact Signatures over NTRU (Falcon)\cite{falcon2020spec}, which contain smaller signatures than ML-DSA. However, Falcon relies on high-precision floating-point arithmetic, making it prone to timing side-channel vulnerabilities that complicate secure real-world implementation. As previously mentioned, hybrid PQC algorithms, which combine pure PQC with traditional algorithms, typically add ~1KB to certificate size. In our paper, we focus primarily on ML-DSA due to its balance of signature size and performance, and SLH-DSA to act as an edge case to examine the impact of larger certificate sizes.

\subsubsection{Merkle Trees}
\label{subsec:merkle}
For TLS 1.3 X.509 certificates, digital signatures are generated by signing a hash of the message using the signer's private key and verifying it with the corresponding public key. Each certificate is required to be signed by its issuer, scaling the certificate size with the chain depth, significantly increasing overall handshake size. Merkle Tree Certificate schemes eliminate intermediate certificates by using a Merkle proof to prove that it is within a CA authority-signed set. We provide a demonstration of the structure of a Merkle Certificate Scheme in Figure~\ref{fig:merkle}. It is important to note that Merkle Tree Certificates are a method of proving certificate authenticity and not actual certificates themselves. They differ from the standard X.509 structure, having the CA authority sign the Merkle root, with each leaf containing a Merkle proof that proves its identity through its hash path to the root, removing the need for certificate chain intermediates. The proof size of an MTC is proportional to the product of the tree height and the hash size\cite{ietf_mtc_2026}. The hash size is 32 bytes when using the SHA-256 algorithm, and the tree height is $log_2(N)$ where N is the number of leaves. (35) Typically, a CA contains anywhere from $2^{24}$ - $2^{28}$ leaves, producing MTC proof sizes of around ~700–900 bytes\cite{cloudflare2025bootstrapmtc}. The MTC leaf size can be expressed as the sum of the original certificate size and the Merkle proof size. While we do not expect much of a decrease in overhead for traditional schemes whose intermediate certificates are not much larger than the proof size, PQC certificates benefit greatly, removing the size complexity introduced by PKI.

\subsection{Content Delivery Networks}
Content Delivery Networks are a system of geographically arranged servers that use caching to reduce content delivery time to users. They implement edge server infrastructure, routing users to the nearest server, focusing on optimizing performance, availability, and scalability to provide users the minimum latency possible. CDNs typically terminate the TLS connection at the edge server, resulting in user connections being between the CDN rather than with the origin server\cite{nygren2010akamai,pathan2008cdn}. The CDN controls all aspects of the handshake, making handshake efficiency critical to overall CDN performance. Larger key and certificate sizes pose a particular challenge for CDNs, which are centered around low latencies. The increased handshake sizes are more sensitive to congestion window limits than classical alternatives, where each additional RTT incurred carries significant latency penalties at CDN scale. Even small amounts of packet loss can significantly affect aggregate latency. CDNs utilize various optimizations such as geographical location, certificate chain compression, and session resumption to lower latency for users. 

\begin{figure}[b]
    \centering
    \includegraphics[width=1.0\linewidth]{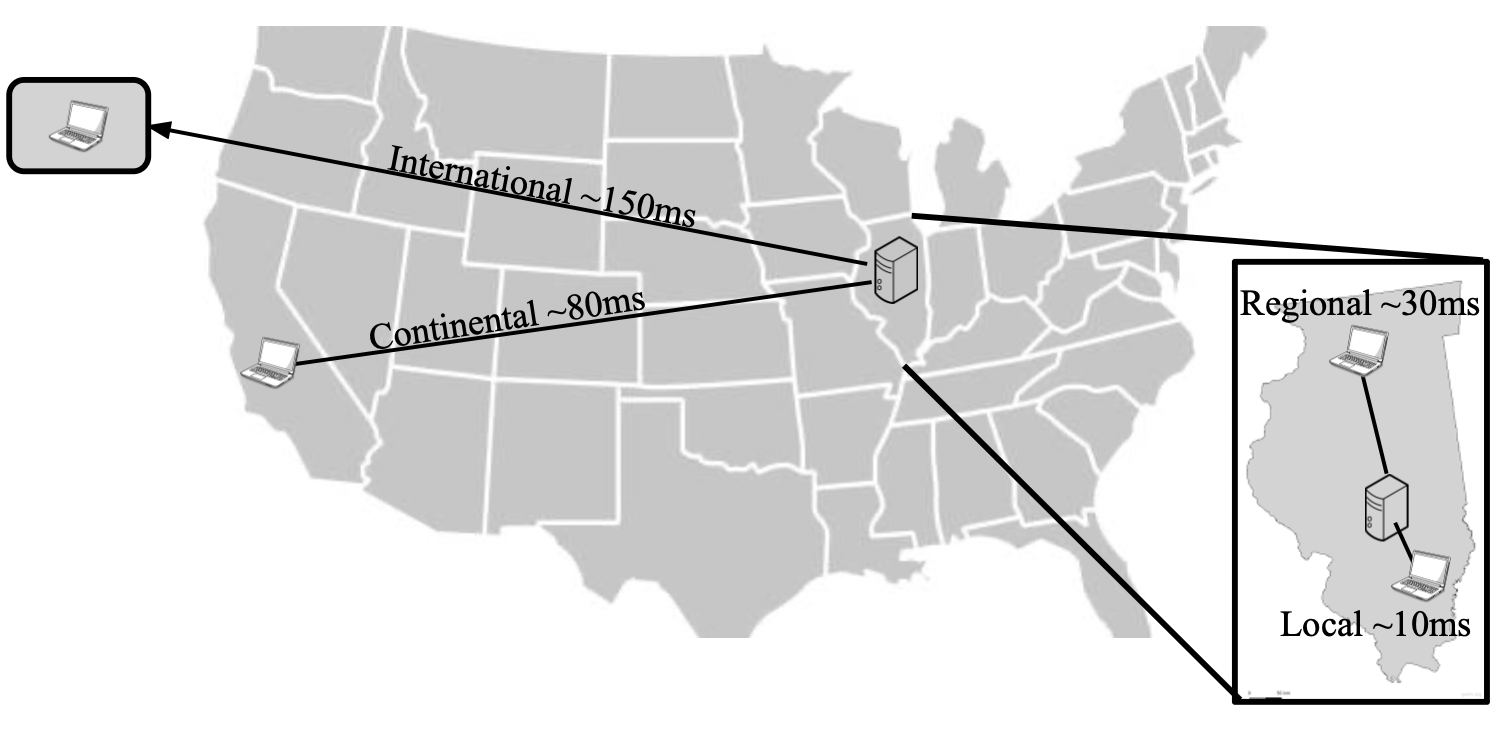}
    \caption{Example Visual of Distances}
    \label{fig:distVis}
\end{figure}

\subsubsection{Geographical Proximity} 
\label{subsubsec:geographical}
RTT, which typically dominates end-to-end latency, is affected by a variety of factors, such as routing inefficiency, network congestion, processing delay, and other overhead. A larger physical distance results in a longer propagation delay, which in turn leads to a higher RTT. Local connections typically exhibit RTTs of ~1-10ms, regional connections ~10-40ms, continental connections ~40-90ms, and intercontinental connections ~120-200+ms\cite{martinez2023rtt}. CDNs are strategically positioned so that users are usually within 10-50ms of a CDN node. Due to their geographical proximity to end users, CDNs are expected to significantly reduce end-to-end latency compared to non-CDN deployments. Since TLS handshakes require at least one round trip, any additional round trips that may occur due to bandwidth penalties increase the latency proportionally\cite{grigorik2013tls}. Additionally, PQC, which has a larger handshake size and is therefore more prone to packet loss and retransmissions, will benefit from a shorter RTT during its increased retransmissions. 

\subsubsection{Certificate Size Optimizations} 
CDNs optimize certificate chains by omitting unnecessary intermediates, choosing shorter trust paths, or using a chain based on the client's root trust CAs. The CDN server is not required to send every intermediate certificate, often sending less if the user already has certificates cached. CDNs also typically maintain multiple pre-configured certificate chains for a domain, targeting different client capabilities (e.g., legacy vs modern browsers), dynamically selecting the shortest compatible chain for each connection\cite{rfc4158}. A CDN's distributed edge architecture allows it to dynamically control which intermediate certificates are served, without being constrained by the fixed configurations typical of origin servers. Although these optimizations are small for lower RTT, they reduce packet loss and provide potential for lowering handshake sizes below window size limits\cite{kozlov2024letsencryptchain}.

\begin{table*}[t]
\centering
\caption{Comparison of Prior Research vs. Our Work}
\label{tab:related_work_comparison}
\renewcommand{\arraystretch}{1.3}
\begin{tabularx}{\textwidth}{@{}l X X@{}}
\toprule
\textbf{Feature} & \textbf{Related Work} & \textbf{This Work} \\
\midrule
\textbf{Primary Variable} & General handshake size and protocol-level overhead. \cite{sikeridis2020overhead,sikeridis2020pqtls} & Isolates \textbf{certificate chain size} as the primary determinant of \textbf{time to first byte}. \\
\textbf{Network Analysis} & Continuous modeling of packet loss and propagation delay. \cite{sikeridis2020overhead,sikeridis2020pqtls} & Identifies discrete \textbf{transport layer flight limits} for PQC (thresholds at 10KB and 40KB). \\
\textbf{Benchmarking} & Direct comparison of pure PQC vs. classical algorithms. \cite{montenegro2024qrtls,montenegro2025pqcnetwork} & Employs \textbf{size-matched ECDSA} simulations to isolate implementation vs. network overhead. \\
\textbf{Optimization} & MTC optimization on PQC certificates and CDN optimizations for classical schemes\cite{ietf_mtc_2026, google2026quantumsafehttps,rfc4158}. & Experimental evaluation of \textbf{MTC and CDN-specific optimizations} for PQC environments. \\
\textbf{Infrastructure} & Generic TLS 1.3 environments and PQC protocol adoption rates. \cite{sowa2024pqcinstrument,dam2023pqcsurvey}& Focuses on \textbf{CDN vs. Non-CDN} performance in latency-sensitive architectures. \\
\textbf{Empirical Grounding} & General internet-wide scanning (e.g., Censys, Zeek). \cite{sowa2024pqcinstrument, 10821264}& Utilizes \textbf{NCSA-monitored TLS traffic} to quantify real-world CDN session resumption rates. \\
\bottomrule
\end{tabularx}
\end{table*}

\subsubsection{Session Resumption} 
Session resumption allows clients to reconnect without requiring a full handshake through the use of session tickets to resume data exchange\cite{rfc5077}. These session tickets eliminate the need for certificate chains, certificate verification, handshake signatures, or full key exchange setup, effectively removing PQC overhead\cite{rfc8446}. While also present in origin servers, it is much more common in CDNs due to their aggressive optimization tendencies. As a result, certificate size impacts on latency may be additionally reduced for CDN structures. Through optimizations, the effect of certificate size on latency has the potential to be reduced using CDN-based architecture.

\section{Related Works}
Post-Quantum deployments with respect to transport security have been widely studied. Particularly, handshake latency and size inflation across PQC-based algorithms in TLS 1.3 have been comprehensively reviewed in works by Sikeridis et al., comparing various certificate schemes under realistic conditions \cite{sikeridis2020pqtls, 10821264}. Subsequent studies have further explored this topic, analyzing key exchanges and authentications showing increases in TLS handshake latencies, specifically measuring how window sizes affect slowdowns\cite{sikeridis2020overhead}.

\textbf{Recent Works.} A high-level summary of topics covered by related work is provided in Table~\ref{tab:related_work_comparison}.  PQC assessments using OQS show that handshake size inflation, cryptographic computation, and network conditions significantly affect latency \cite{montenegro2024qrtls, montenegro2025pqcnetwork}, with bandwidth and fragmentation dependencies noted across key exchange schemes \cite{sikeridis2020overhead}. Although certificate hierarchy work has examined chain configurations and placement under PQC TLS 1.3 \cite{sikeridis2020pqtls}, a key gap remains in that no prior work isolates certificate chain size as an independent latency contributor.

Prior work on transport layer evaluations has examined TCP congestion control, MTU limits, and initial window sizes as determinants of handshake overhead, including how larger handshake sizes can trigger additional round trips due to fragmentation. However, these factors are typically measured in combination, rather than isolating the discrete latency thresholds with respect to the size of the certificate chain under PQC \cite{sikeridis2020overhead,sikeridis2020pqtls}.

CDNs have been widely studied for latency-reduction strategies such as session resumption and edge server placement, which reduce overhead by eliminating certificate retransmission and minimizing RTT \cite{dean2013tail}. Certificate chain size reductions have also been examined as a mitigation strategy \cite{kozlov2024letsencryptchain}, though these optimizations are largely limited to traditional schemes with minimal PQC exploration. No existing work directly compares CDN-specific certificate chain optimizations under PQC, nor examines how such optimizations relate to transport layer bandwidth thresholds. 

MTC reduces certificate transmission by replacing X.509 chains with a single proof, though prior work remains focused on design and implementation rather than empirical evaluation under realistic conditions\cite{google2026quantumsafehttps}.

Internet-wide tools like Zeek have been used to analyze TLS deployments \cite{10821264}, but lack direct CDN vs. non-CDN comparisons and connections to PQC simulations.

\textbf{Novelty.} Unlike previous work, this paper isolates certificate chain size as the primary variable of study, constructing PQC certificate chains of varying sizes and mirroring these sizes in traditional X.509 simulations for comparison. The contrast of the two allows us to identify how latency varies between the two as certificate chain sizes increase. We also analyze and indicate where transport-level packet flight limits occur, noting their discrete increases in TTFB with respect to certificate size. We propose the usage of Merkle Tree Certificates (MTC) and CDN chain size optimizations, namely session resumption and certificate size optimizations, to keep handshake sizes below packet flight thresholds. We additionally quantify the ranges and savings of staying below the said threshold for our simulated data.


\section{Our CDN Findings}
We analyze traffic from an NCSA with national traffic data to evaluate real-world TLS deployments and session resumption rates for CDN and non-CDN servers. Our data set comprises Zeek-monitored \cite{zeek_ssl_main,paxson1998bro} TLS logs from real-world traffic that include metadata on conducted handshakes, such as protocol version, session reuse indicators, and other certificate-related fields. While the logs contained various data about certificates and handshakes, we mainly analyzed the TLS adoption and session resumption rates.

We classified CDNs based on their Autonomous System Number (ASN), using GeoIP's IP to ASN mapping to associate each entry with an organization\cite{maxmind_geolite2}. We compared the ASN to a list of known CDN ASNs to label CDN endpoints. Note that our CDN ASN label was strict in order to specifically observe CDN behaviors. We classified our data into 4 groups, with the groups being: 1.) CDN, 2.) Cloud, 3.) Non-CDN, and 4.) unidentified. We intentionally separated them into distinct categories, only comparing CDN and non-CDN cases.

\textbf{Observed Trends.} Table~\ref{tab:tls_resumption_cdn} shows that CDNs have a significantly higher TLS 1.3 adoption rate (84.74\%) compared to non-CDNs (75.73\%), consistent with CDNs' tendency to prioritize modern, low-latency protocol configurations. More importantly, session resumption rates are much higher for CDNs, with 94.16\% of TLS 1.3 connections using session resumption as compared to 46.09\% of non-CDNs. Overall, CDNs had an 80.30\% resumption rate, whereas non-CDNs only had 35.44\%. We note that most modern systems, as determined by their usage of TLS 1.3, use session resumptions with CDNs, and legacy systems most likely do not support session resumption, bringing down the rates.

\begin{figure}[t]
    \centering
    \includegraphics[width=1.0\linewidth]{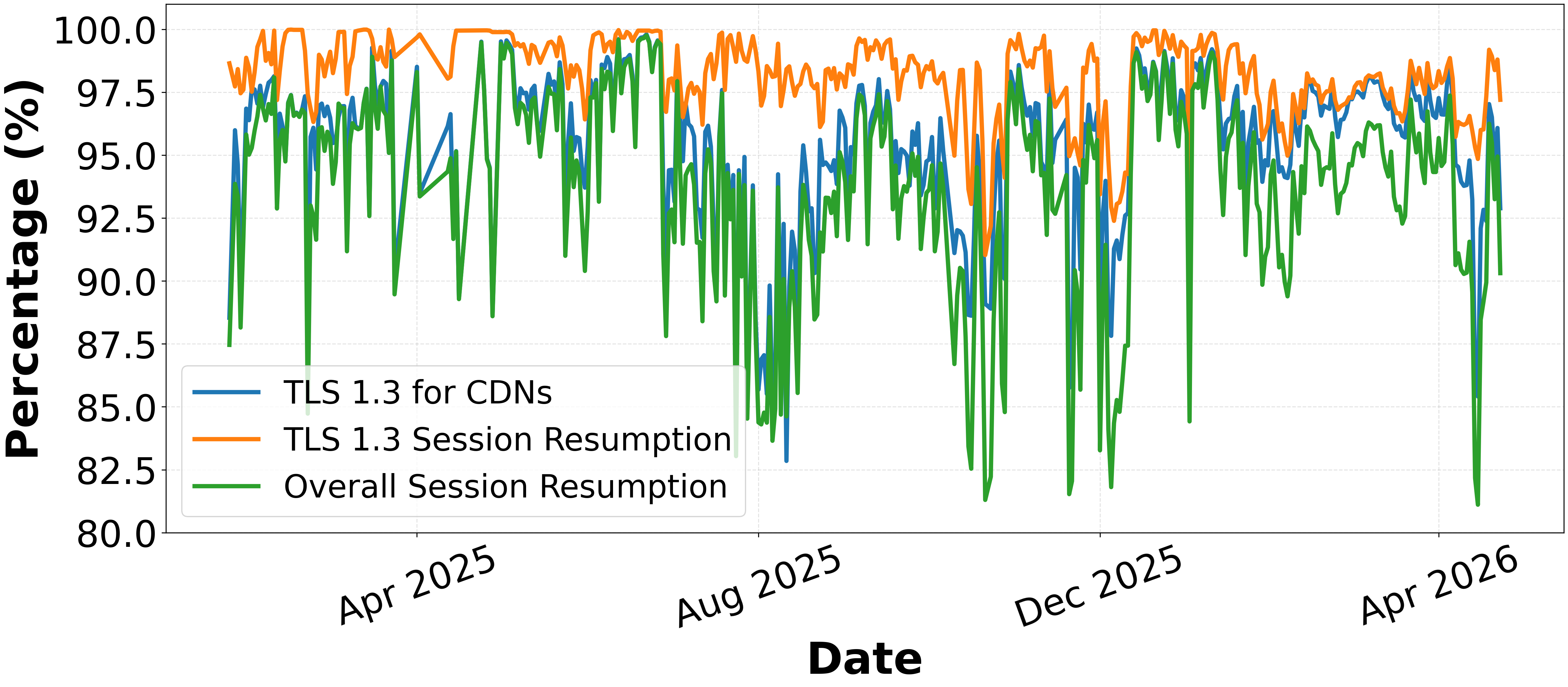}
    \caption{Session Resumption and TLS 1.3 Rates from 16 months of NCSA data.}
    \vspace{-4mm}
    \label{fig:allData}
\end{figure}

We also provide a measurement of the trend of session resumption since January 2025 until the time of writing this paper, April 2026 using the same NCSA data in Figure~\ref{fig:allData}. The plot supports our conclusion as we see correlation between the TLS 1.3 and overall session resumption rate. We thus see that as TLS 1.3 continues to become more prevalent, session resumption rates will likely also rise.

However, we have to note several limitations of our setup. First, ASN-based classification is not fully accurate, as cloud providers can be both CDN or non-CDN, causing possible mis-classification. Additionally, Zeek logs only provide metadata rather than full packet captures, which does not allow full visibility into fragmentation, packet loss, or precise timings. More detailed data would allow controlled experiments of TTFB in real settings, allowing for a comparison of TTFB for resumed handshakes. 

\textbf{Empirical Insight.} Despite limitations, this is the first paper to perform measurements of certificate chain size on TTFB. These results allow us to see real-world applications of session resumption and how CDNs aggressively optimize their connections to reduce handshake overhead. This is particularly important in the context of PQC, where session resumption may allow for reduced latencies of larger certificate chains in realistic deployments.

\begin{table}[t]
\centering
\caption{Session Resumption Rates for CDN and Non-CDN Endpoints}
\label{tab:tls_resumption_cdn}

\renewcommand{\arraystretch}{1.2}
\setlength{\tabcolsep}{6pt}

\begin{threeparttable}
\begin{tabularx}{\columnwidth}{Xcc}
\toprule
\textbf{Metric} & \textbf{CDN} & \textbf{Non-CDN} \\
\midrule
TLS 1.3 Adoption (\%) & 84.74 & 75.73 \\
\midrule
Session Resumption (TLS 1.3, \%) & 94.16 & 46.09 \\
Session Resumption (All, \%) & 80.30 & 35.44 \\
\bottomrule
\end{tabularx}

\begin{tablenotes}
\footnotesize
\item Data collected over a 24-hour snapshot from an NCSA. (Apr. 14 2026)
\end{tablenotes}
\end{threeparttable}
\end{table}

\begin{table}[b]
\centering
\begin{threeparttable}
\caption{Comparison of Certificate Component Sizes Across Schemes}
\label{tab:cert_sizes}
\renewcommand{\arraystretch}{1.2}
\begin{tabularx}{\columnwidth}{l X cc}
\toprule
\textbf{Scheme} & \textbf{Configuration} & \textbf{Leaf (KB)} & \textbf{Intermediate (KB)} \\
\midrule
\multirow{1}{*}{ECDSA}
 & Standard Chain & 1.0 & 2.0 \\
\midrule
\multirow{2}{*}{SLH-DSA}
 & Standard Chain & 16.6 & 32.1 \\
 & MTC Variant    & 17.6 & --   \\
\midrule
\multirow{2}{*}{ML-DSA}
 & Standard Chain & 3.9 & 8.0 \\
 & MTC Variant    & 4.8 & --   \\
\bottomrule
\end{tabularx}
\begin{tablenotes}
\footnotesize
\item MTC entries omit intermediate values, as the scheme does not utilize intermediate certificates.
\end{tablenotes}
\end{threeparttable}
\end{table}

\section{Our PQC Simulations Testbed}
\label{sec:CDNSim}
This section is focused on isolating and analyzing how latencies vary across different certificate chain schemes and sizes. We compare how Post-Quantum implementations of increased certificate chains compare to similarly sized traditional certificate chains, specifically noting how size increase impacts latency, measuring the time to first byte (TTFB) as our main metric. For our simulations, we use OpenSSL, performing a TLS handshake and HTTP GET request, to determine the time from sending the request until the first byte received. We chose this as our primary metric as it has not been widely studied and it reflects the most realistic real world usage, directly measuring the time until a response.

\textbf{Experimental Setup.} For all of our measurements, we ran a simple OpenSSL TLS s\_server, sending an HTTPS request through that TLS connection. For communication between our client and server, we used two Amazon Web Services EC2 Ohio servers to simulate real network behaviors, keeping both client and server in the same region so as not to introduce too many external variables. To introduce delay into our simulations, we used \textit{tc netem}, a traffic control network emulator, injecting delay on both the forwards and reverse path. It is important to note for our values, we include time of start up overhead, measuring the full end-to-end execution time rather than just the time corresponding to the raw network data.

To emulate varying certificate chains for Post-Quantum as well as traditional infrastructure, we artificially inflated the certificate sizes by adding non-critical certificate extensions so as not to affect certificate verification time, padding the DER encoded length of the certificate. Thus, we were able to solely analyze the network overhead due to increased certificate chain size without affecting certificate verification times. We were able to perform this for Post-Quantum certificates in addition to traditional ones through Open Quantum Safe's (OQS) open source library. (33) It is important to note that at the time, AWS didn't support OpenSSL 3.5, which is required for OQS, causing us to run the servers on Docker containers. We compare the inflated traditional certificates to inflated Post-Quantum certificates to note whether the difference in TLS stacks had any effect on latencies as certificate chain sizes increased.

\begin{figure}[b]
    \centering
    \includegraphics[width=1.0\linewidth]{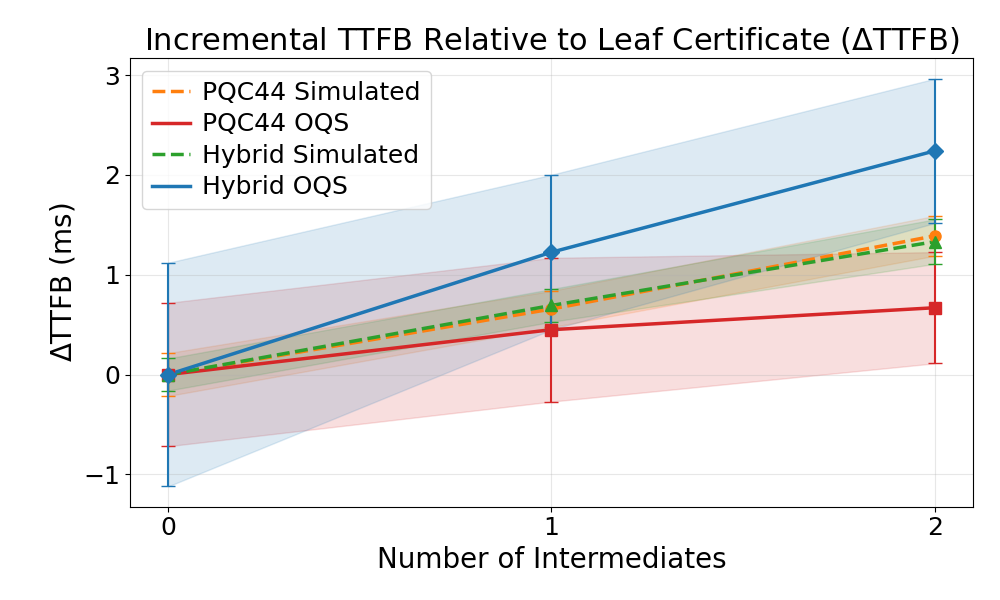}
    \caption{Minimal OQS vs TLS implementations}
    \label{fig:oqs}
\end{figure}

\begin{table}[b]
\centering
\begin{threeparttable}
\caption{Comparison of TTFB Between Classical Size-Matched Simulations and PQC (OQS)}
\label{tab:oqs_vs_sim}
\renewcommand{\arraystretch}{1.2}
\begin{tabularx}{\columnwidth}{l X cc}
\toprule
\textbf{Scheme} & \textbf{Certificate Chain} & \textbf{Simulated (ms)} & \textbf{OQS (ms)} \\
\midrule
\multirow{3}{*}{ML-DSA}
 & Leaf (3.9 KB) & $5.46 \pm 0.22$ & $56.81 \pm 0.72$ \\
 & Leaf + 1 Int. (7.9 KB) & $6.11 \pm 0.18$ & $57.26 \pm 0.72$ \\
 & Leaf + 2 Ints. (11.9 KB) & $6.85 \pm 0.20$ & $57.48 \pm 0.56$ \\
\midrule
\multirow{3}{*}{Hybrid ML-DSA}
 & Leaf (4.8 KB) & $5.47 \pm 0.16$ & $60.79 \pm 1.12$ \\
 & Leaf + 1 Int. (9.5 KB) & $6.16 \pm 0.16$ & $62.01 \pm 0.78$ \\
 & Leaf + 2 Ints. (14.3 KB) & $6.80 \pm 0.22$ & $63.03 \pm 0.72$ \\
\bottomrule
\end{tabularx}
\begin{tablenotes}
\footnotesize
\item While increasing the number of intermediates impacts ECDSA size-matched certificates, propagation delay effects are limited in the OQS implementation.
\end{tablenotes}
\end{threeparttable}
\end{table}
\subsection{Evaluation of Open Quantum Safe Distinctions}
\label{subsec:OQSComp}
We begin our study by comparing the OQS implementation TTFB to a standard TLS 1.3 stack to assess whether the differing TLS stacks cause any disparities. We start with the minimal handshake, measuring the TTFB for both cases. We ran 100 simulations for each case with this setup, computing the mean and standard deviation for each one. The time to first byte was measured by starting a timer, establishing a TCP connection, completing the full TLS handshake, sending an HTTPS request, and stopping the timer after receiving the first byte. We can model the TTFB as follows:
\begin{equation}
\mathrm{TTFB} = T_{\mathrm{TCP}} + T_{\mathrm{TLS}} + T_{\mathrm{request}} + T_{\mathrm{response}}
\end{equation}
\textbf{Certificate Configuration.} To compare only the difference between the PQC and traditional TLS stack, we ran a minimal handshake. We only measured the TTFB of a raw TLS connection without enforcing certificate verification and sending a minimal HTTP request. We carried this out for ML-DSA and Hybrid ML-DSA, using OQS first to run the minimal session, and then padding traditional ECDSA certificate sizes to size match the ML-DSA certificates. We used ML-DSA-44 and P256-ML-DSA-44 for our specific PQC certificates and P256 for our specific ECDSA certificate given their widespread usage. We also used the hybrid ML-KEM-768 key. To fully isolate how certificate chain size affects the TTFB, we experimented with having 0, 1, and 2 intermediate certificates, observing how the TTFB grew. 

\textbf{Limited Impact of Propagation Delay.} We observe that from our results in Table~\ref{tab:oqs_vs_sim} that OQS has a significantly larger TTFB than our simulated data, with a roughly 50-55ms difference. This difference was likely due to PQC key and certificate generation overhead or the TLS stack difference of OQS. Since the minimal TLS connection produced a handshake of roughly the same size, the difference is likely attributable to the TLS stack itself rather than the certificate chain size. To show this, we graphed the incremental TTFB relative to the number of intermediate certificates added, as shown in Figure~\ref{fig:oqs}. We see that as the number of intermediates, and therefore certificate chain sizes increase, Post-Quantum and traditional schemes remain relatively similar in incremental TTFB. As the certificate chain size increases, the increase in TTFB is nearly negligible compared to the total TTFB increase, suggesting that \textit{for PQC the TTFB is dominated by implementation rather than actual network propagation}. We conclude that under our tested conditions, end-to-end latency is dominated by TLS and protocol differences rather than increased latency due to size inflation.

The OQS results for hybrid being an additional $\sim$4ms cannot be explained by certificate size alone, as hybrid chains are only $\sim$1 KB larger, and our experiments show minimal TTFB differences between increased certificate-sized schemes. This indicates that the observed increase is most likely due to implementation overhead rather than transmission cost. The similar incremental delays observed with each additional certificate indicate that the impact of certificate size is largely independent of whether we use inflated ECDSA or ML-DSA certificates. These results suggest that our simulations of artificially inflated certificates are sufficiently representative of the effects of modeling TTFB for certificate chains increase. Our results also indicate that TTFB is not primarily driven by propogation delay in OQS systems, implying in real world deployments, latency is instead largely influenced by overhead from differences in implementation.

\begin{table*}[t]
\centering
\begin{threeparttable}
\caption{Time to First Byte Across Round Trip Times for Different Certificate Schemes (ECDSA Based)}
\label{tab:ttfb_rtt_ecdsa}
\renewcommand{\arraystretch}{1.2}
\begin{tabularx}{\textwidth}{X ccccc}
\toprule
\textbf{Certificate Scheme} 
& \textbf{No RTT } 
& \textbf{10 ms } 
& \textbf{50 ms } 
& \textbf{100 ms } 
& \textbf{200 ms } \\
\midrule
ECDSA X.509      
& $8.06 \pm 0.31$ & $28.71 \pm 0.17$ & $109.00 \pm 1.18$ & $208.84 \pm 0.20$ & $409.10 \pm 1.80$ \\
ML-DSA           
& $7.98 \pm 0.37$ & $28.88 \pm 1.65$ & $108.77 \pm 0.19$ & $208.80 \pm 0.22$ & $408.83 \pm 0.25$ \\
\textbf{SLH-DSA}          
& $\mathbf{8.29 \pm 0.23}$ & $\mathbf{39.20 \pm 0.20}$ & $\mathbf{159.41 \pm 0.32}$ & $\mathbf{309.41 \pm 0.34}$ & $\mathbf{609.45 \pm 0.57}$ \\
MTC + ML-DSA     
& $7.47 \pm 0.28$ & $28.28 \pm 0.18$ & $108.96 \pm 0.95$ & $208.37 \pm 0.16$ & $408.63 \pm 1.28$ \\
\textbf{MTC + SLH-DSA}    
& $\mathbf{7.73 \pm 0.34}$ & $\mathbf{40.03 \pm 0.90}$ & $\mathbf{158.77 \pm 0.18}$ & $\mathbf{309.20 \pm 1.38}$ & $\mathbf{608.84 \pm 0.24}$ \\
\midrule
Session Resumption 
& $7.01 \pm 0.14 $ & $27.39 \pm 0.15 $ & $107.49 \pm 0.15$ & $207.60 \pm 1.00$ & $407.63 \pm 1.01$ \\
\bottomrule
\end{tabularx}
\begin{tablenotes}
\footnotesize
\item Note: Each entry is the TTFB 
\end{tablenotes}
\end{threeparttable}
\end{table*}

\begin{table*}[t]
\centering
\begin{threeparttable}
\caption{Time to First Byte Across Round Trip Times for Different Certificate Schemes (ML-DSA Based)}
\label{tab:ttfb_rtt_ML-DSA}
\renewcommand{\arraystretch}{1.2}
\begin{tabularx}{\textwidth}{X ccccc}
\toprule
\textbf{Certificate Scheme} 
& \textbf{No RTT } 
& \textbf{10 ms } 
& \textbf{50 ms } 
& \textbf{100 ms } 
& \textbf{200 ms } \\
\midrule
ML-DSA           
& $331.08 \pm 7.53$ & $381.50 \pm 38.93$ & $538.21 \pm 10.08$ & $748.01 \pm 49.55$ & $1155.84 \pm 54.33$ \\
\textbf{SLH-DSA}          
& $\mathbf{341.43 \pm 47.79}$ & $\mathbf{385.37 \pm 8.54}$ & $\mathbf{594.48 \pm 29.44}$ & $\mathbf{845.59 \pm 8.80}$ & $\mathbf{1357.31 \pm 55.95}$ \\
MTC + ML-DSA     
& $333.04 \pm 24.55$ & $370.35 \pm 7.46$ & $544.16 \pm 72.77$ & $743.22 \pm 36.80$ & $1157.77 \pm 83.97$ \\
\textbf{MTC + SLH-DSA}    
& $\mathbf{328.96 \pm 7.35}$ & $\mathbf{392.08 \pm 78.71}$ & $\mathbf{591.42 \pm 33.09}$ & $\mathbf{849.04 \pm 57.45}$ & $\mathbf{1351.17 \pm 31.35}$ \\
\midrule
Session Resumption 
& $340.19 \pm 24.35$ & $381.35 \pm 23.49$ & $547.51 \pm 30.79$ & $740.31 \pm 10.39$ & $1152.69 \pm 41.05$ \\
\bottomrule
\end{tabularx}
\begin{tablenotes}
\footnotesize
\item Note: Each entry is the TTFB (ms)
\end{tablenotes}
\end{threeparttable}
\end{table*}

\subsection{RTT Impact on TTFB Across Varying Certificate Schemes}
\label{subsec:session_resumption}
In this section, we aim to compare how TTFB is affected by varying certificate chain sizes for different RTTs. We not only use the RTTs to analyze the effect of transmission delays, but to also mimic differing geographic distances. 

We use the same setup as described in Section~\ref{sec:CDNSim}, again running 100 simulations for each case, measuring mean and standard deviations. However, for this evaluation, we introduce delay into our simulations by injecting a 2-way RTT. In order to further explore more realistic conditions, we opted to use a full TLS handshake with certificate verification along with the HTTP request. In addition to our varying certificates, we include a session resumption case for each certificate scheme, where we measure the TTFB using session tickets rather than a full handshake. For the resumed session, we first ran a full TLS handshake to establish the session and then measured the subsequent connections of that session. 
To examine how TTFB changed with varying certificate chain sizes, we used chain sizes corresponding to real certificate schemes, with chain size varying as a function of the cryptographic scheme. It is important to note that we didn't implement the actual cryptographic protocols or Merkle Tree structure, only artificially simulating their sizes by padding with certificate extensions. Additionally, we only used 1 intermediate to not introduce too many variables. For choosing the size of certificates,  we refer to NIST standards, specifically looking at ML-DSA-44 and SLH-DSA-192s. (34)(35) ML-DSA-44 serves as a representative baseline for post-quantum signature schemes with moderate size overhead, while SLH-DSA-192s was chosen due to its larger certificate size to act as an edge case. The sizes we used are noted in Table~\ref{tab:cert_sizes}.

\textbf{Geographical Impact.} We ran our experiments for both ECDSA and ML-DSA-based certificates as seen in  Table~\ref{tab:ttfb_rtt_ML-DSA} and Table~\ref{tab:ttfb_rtt_ecdsa}. We use RTT to approximate the geographical distances as referenced in  Figure~\ref{fig:distVis}. While we see a strong impact of RTT for our ECDSA-based simulation, with RTT dominating the TTFB, the ML-DSA simulations remain less affected. We still see that ML-DSA simulations are largely influenced by RTT at higher RTTs, but in most cases, and in cases of CDNs, RTT will rarely reach such edge cases. As discussed previously, the large overhead of Post-Quantum Cryptography has a substantial impact on latencies, especially for CDNs.

\textbf{ML-DSA vs. ECDSA.} We observe that TTFB values remained relatively similar for ECDSA and all ML-DSA cases, as no significant increases in ms were observed. However, for both SLH-DSA and SLH-DSA + MTC cases, we see that the TTFB increases by a whole RTT. Within the ECDSA and ML-DSA cases and the SLH-DSA cases, we are unable to observe a noticeable difference in TTFB due to the reduction in certificate chain size that MTC would be expected to introduce. This can be explained in Table~\ref{tab:oqs_vs_sim}, where we note that increases in TTFB due purely to size were on the scale of tenths of microseconds, which we also noted were overshadowed by the latency increases caused by TLS stack implementations. We note much larger standard deviations for our full experiments due to simulating a full connection in these simulations, masking out the latency impacts of increased certificate chain sizes. 
While we ran a session resumption case for each certificate scheme, we noticed that the data stayed relatively the same across all of our cases. This is consistent with how session resumption functions since it doesn't depend on certificates or keys, instead only verifying the session ticket. One limitation of our simulation of session resumption was that the TTFB values reflect end-to-end application response time rather than just TLS handshake latency. Client and server process startup overhead in addition to TCP connection establishment could have caused the session resumption values to be masked by the larger overhead of other components.
\begin{figure}[t]
\centering

\includegraphics[width=\columnwidth]{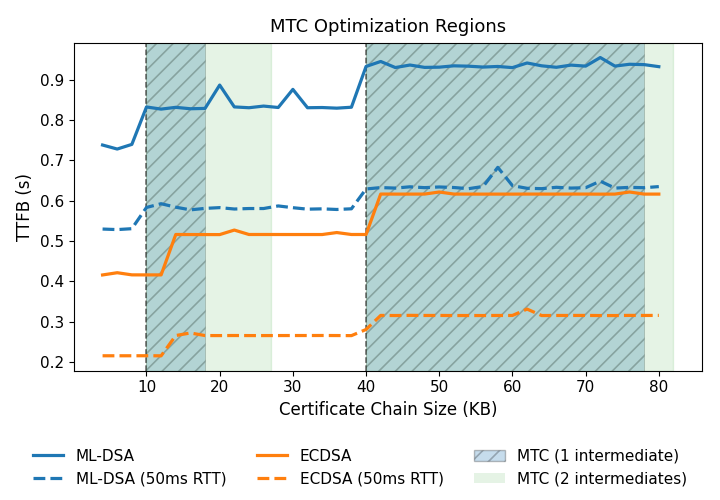}
\caption{Certificate Chain Size Optimizations allowing for bypassing of bandwidth penalty thresholds}
\label{fig:ttfb_vs_size}

\vspace{6pt}

\captionsetup{type=table}
\caption{Optimization Regions Across Certificate Size Thresholds}
\label{tab:optimization_regions}

\begin{minipage}{\columnwidth}
\centering
\renewcommand{\arraystretch}{1.25}
\setlength{\tabcolsep}{8pt}

\begin{tabular*}{\columnwidth}{@{\extracolsep{\fill}} l c c}
\toprule
\textbf{Optimization} & \textbf{Lower Bound (KB)} & \textbf{Upper Bound (KB)} \\
\midrule
MTC (1 int.)  & 10--18  & 40--78  \\
MTC (2 int.)  & 10--27  & 40--117 \\
CDN (25\%)    & 10--13  & 40--53  \\
CDN (40\%)    & 10--17  & 40--67  \\
\bottomrule
\end{tabular*}
\end{minipage}

\end{figure}

\begin{figure}[t]
    \centering
    \includegraphics[width=1.0\linewidth]{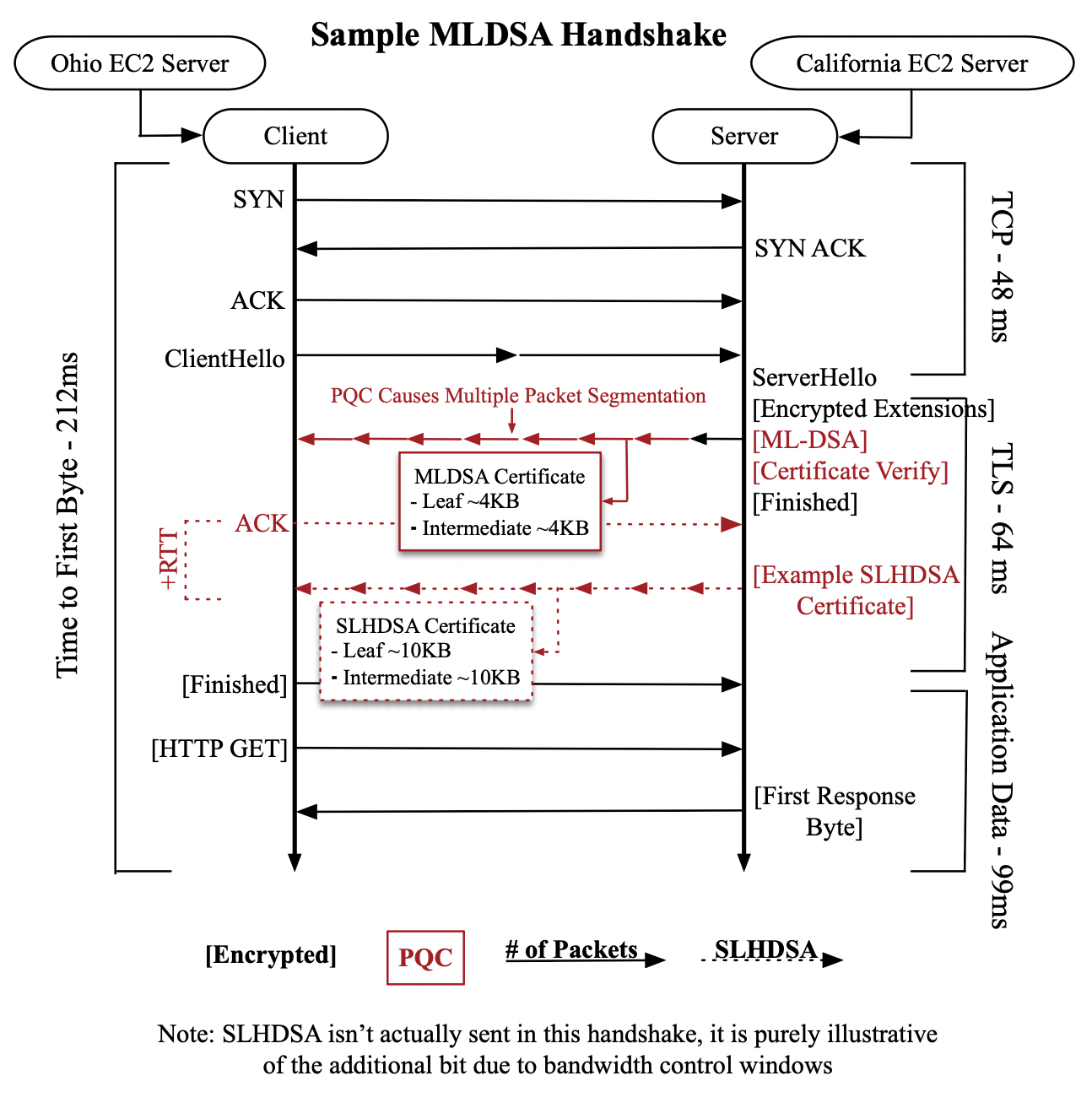}
    \caption{Our Sample ML-DSA Handshake Between Ohio and California EC2 Servers}
    \vspace{-4mm}
    \label{fig:mlHand}
\end{figure}

\textbf{Bandwidth Flight Windows.} The important takeaway of our simulations is the added RTT for the SLH-DSA cases increased TTFB by up to 1.5x for extreme cases. While transmission delay due to size caused increased TTFBs as we saw in Figure~\ref{fig:oqs}, it doesn't cause the addition of a full RTT. The most likely explanation for this is that there are bandwidth limitations on the network, or more specifically, packet transmission windows, which \textit{limit how much data can be "in flight" at once}. From our data, we observe that ML-DSA and ECDSA are small enough to fit within one flight of data, whereas SLH-DSA crosses the flight boundary "threshold". Since the TTFB is dominated by mostly non-network-related latencies, ML-DSA is small enough so that it has similar overhead due to size as traditional X.509 certificates. We provide a figure demonstrating the extra RTT introduced by SLH-DSA as compared to an ML-DSA certificate in Figure~\ref{fig:mlHand}. The figure shows a simulated PQC handshake between a California and Ohio EC2 AWS server using only ML-DSA. The SLH-DSA certificate was not included in the transmission, only acting to demonstrate how exceeding the Bandwidth Initial Window adds an RTT.

For our OQS simulation, we see that for certificate chain sizes within packet transmission window sizes, certificate size has little to no effect on the TTFB. A point of interest, however, is whether the certificate size optimizations are able to reduce the certificate size under the threshold, fitting in a single flight of data. We also note session resumptions being able to achieve the same effect, removing multiple RTTs in extreme cases.

\subsection{Mitigating Bandwidth Penalty Through Certificate Size Optimizations}
We analyze the threshold limits discussed in the previous section, noting precisely where they occur and how certificate chain optimizations are able to lower certificate chains under said thresholds. We maintain the same setup as Section~\ref{subsec:session_resumption}, instead opting to vary certificate size more continuously to isolate where extra RTTs are added. We start with a certificate size of 4KB, increasing by 2KB until 80KB. We used 10 trials per data point and ran simulations for ML-DSA and ECDSA-based algorithms for RTTs of 50 and 100ms. We didn't measure lower RTTs as the OQS implementation obscures any differences they make.

We notice a couple fluctuations in Figure~\ref{fig:ttfb_vs_size} for the values of ML-DSA-based certificates, largely due to network variability being amplified by the RTT. We see our RTT spikes occur at relatively the same certificate chain sizes for ECDSA and ML-DSA, being around 10KB and 40KB. We also observe that RTT doesn't affect where thresholds occur, only affecting the TTFB.

\textbf{Optimization Heuristics.} We approximate the certificate chain savings of Merkle Tree structures as $\frac{\text{ChainSize}}{2} + 1$ KB for 1 intermediate and $\frac{\text{ChainSize}}{3} + 1$ KB for 2 intermediates. Our MTC size were derived from the underlying Merkle Tree structure as demonstrated in Section~\ref{subsec:merkle} rather than arbitrary assumptions. As a result, while we are unable to directly implement MTC, the estimated certificate sizes remain a realistic approximation of network behavior even without full implementation. We are unable to directly simulate CDN chain optimizations, approximating them according to several studies that claim chain sizes can be reduced by up to 40-50\%\cite{rfc8879,letsencrypt2023chain}. While we are unsure whether chain optimizations are applicable to PQC, we evaluate their potential impact on TTFB. Thus, we model savings as $(Chain Size) *0.75$ for moderate reduction and $(Chain Size)*0.60$ for more aggressive optimizations. 

\textbf{MTC and CDN Savings.} We provide the ranges in which MTC and CDN optimizations are able to lower the certificate chain size enough to be under the BDP threshold for our simulated setup in Figure~\ref{fig:ttfb_vs_size}, providing our exact values in Table~\ref{tab:optimization_regions}. Within our simulation data, MTC optimizations enable certificate chains with one intermediate, ranging from 10–18 KB and 40–78 KB, to avoid bandwidth penalties entirely, while two-intermediate chains remain unaffected up to 10–27 KB and 40–117 KB. For CDN optimizations, we observe smaller ranges with chain size extensions from 10 to ~13-17KB and 40 to ~53-67KB. We see that for one intermediate, MTC allows for nearly \textit{double the size of certificate chains}, and for two intermediates, it allows for nearly \textit{triple}. Note that CDN optimizations that we approximate remain unaffected by intermediate count. We see that CDN optimizations allow for a \textit{~130\% to 167\%} increase in certificate chain sizes according to our assumptions.

\begin{figure}[t]
    \centering
    \includegraphics[width=1.0\linewidth]{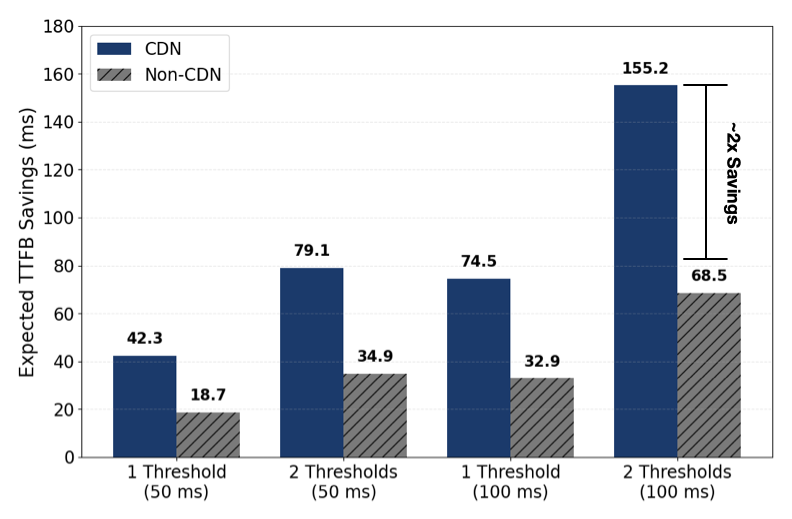}
    \caption{Average savings of $\sim$2x due to session resumption for CDNs compared to non-CDNs}
    \vspace{-2mm}
    \label{fig:sResump_savings}
\end{figure}

\textbf{Session Resumption Savings.} We also demonstrate possible savings due to session resumption bypassing overhead from bandwidth penalties in Figure~\ref{fig:sResump_savings}. Using the data gathered from Zeek, we multiply the total session resumption rates found in Table~\ref{tab:tls_resumption_cdn} with the savings from Figure~\ref{fig:ttfb_vs_size}. We approximate the session resumption TTFB as the TTFB found before the first threshold. For 50ms and 100ms RTTs, we see savings from \textit{~40-80ms} and \textit{~75-155ms}, respectively, depending on the certificate chain size, with nearly twice the TTFB savings in CDN compared to non-CDNs. Moreover, as TLS 1.3 adoption continues to increase, we expect to see higher rates of average savings. While the TTFB savings we found are modest compared to the total TTFB, small differences in latency can accumulate at scale, especially for CDNs. 

\subsection{Limitations}
We note several limitations in our setup. Although our OQS simulations allowed for experimentation with PQC certificates, we had limited insight into applications across various TLS stacks. Furthermore, we were unable to test real applications of Merkle Tree-configured certificates, nor meaningfully reduce certificate chain sizes with CDN optimizations. Additionally, our measurements were conducted without specifically analyzing packet loss or failure rates, which may affect TTFB in real deployments. The precision of our ASN classification could be improved to provide more accurate results.  Despite these limitations, our primary conclusions were unlikely to be affected by these small variations.

\section{Post-Quantum Certificate Deployment Implications}
Our results show that the impact of certificate chain size is mainly caused by network-level constraints rather than by transmission overhead alone. In our experiments, we observe clear bandwidth window thresholds, specifically at 10 and 40 KB-sized certificate chains for our Post-Quantum implementation. Crossing these limits introduces an additional RTT to the total TTFB, with the TTFB jumping discretely at these values. This indicates that TTFB remains relatively the same between threshold ranges, increasing disproportionately across those boundaries.

\textbf{Key Insight 1.} \textit{End-to-end delay is largely dominated by bandwidth limits and implementation overhead rather than pure packet propagation.} In comparing padded ECDSA certificates to their ML-DSA counterparts, we isolated network and implementation effects, allowing us to directly analyze the impact of certificate chain sizes on latency. We saw that the OQS measurements were larger than the traditional implementations of TLS 1.3, indicating that in practical deployments, TTFB is dominated by implementation rather than data transfer overhead. Moreover, by mimicking the behavior of larger geographic distances with an increased RTT, we observe that packet transmission delays negligibly affect TTFB compared to bandwidth packet window restrictions. 

\textbf{Key Insight 2.} \textit{Certificate chain optimizations are able to reduce TTFB by keeping chain size within the bandwidth delay threshold.} We compare approximated MTC and CDN optimizations to reduce certificate chain size to be under bandwidth limit penalties. We observe $\sim$2-3x-sized certificates with MTC and up to $\sim$1.6x certificate chains with CDNs while still remaining below thresholds. The minimal TTFB impact of larger certificates indicates that implementations can adopt more secure or robust certificate chains without incurring substantial additional overhead.

\textbf{Key Insight 3.} \textit{NCSA-backed CDN measurements of session resumption provides additional insight for maintaining TTFB despite bandwidth limits.} Large TTFBs have consequences exemplified in CDNs, which are centered around low-latency connections. We see that session resumption is also able to bypass the aforementioned boundaries, saving RTT. We integrate our findings with Zeek-monitored NCSA data to compare session resumption rates and savings across CDNs and non CDNs. Additionally, we analyze trends over a year-long period to demonstrate session resumption rates over time, noting a correlation between TLS 1.3 and overall session resumption rates.


\section{Conclusion and Future Works}
We discuss various implementations of Post-Quantum Certificates across various schemes and structures, specifically focusing on certificate chain size impact on TTFB. We note the predominant cause of TTFB increase and present several ways to reduce it, supporting our results with real-world data from an NCSA and discuss implications for future PQC implementation.

Some directions of future work are to examine fragmentation and packet loss for varying certificate chain sizes, particularly noting how certificate chain size reductions would benefit latencies in Post-Quantum applications. Another direction of future study is using more detailed data to better model the effects of CDN properties on latency and running certificate chain simulations on real-world CDN infrastructure.



\bibliographystyle{IEEEtran}
\bibliography{breferences}

\end{document}